# Digital Ecosystems: Enabling Collaboration in a Fragmented World

*Short Paper*

**Marc Schmitt**
University of Oxford
marc.schmitt@cs.ox.ac.uk

## Abstract

*As geopolitical, organizational, and technological fragmentation deepens, resilient digital collaboration becomes imperative. This paper develops a spectrum framework of polycentric digital ecosystems—nested socio-technical systems spanning personal, organizational, inter-organizational, and global layers. Integration across these layers is enabled by four technology clusters: AI and automation, blockchain trust, federated data spaces, and immersive technologies. By redefining digital ecosystems as distributed, adaptive networks of loosely coupled actors, this study outlines new pathways for cross-border coordination and innovation. The framework extends platform theory by introducing a multi-layer conceptualization of polycentric digital ecosystems and demonstrates how AI-enabled infrastructures can be orchestrated to achieve digital integration in a fragmented, multipolar world.*

## Introduction

The era of seamless globalization is fading. In recent years, geopolitical fragmentation has intensified as shifting alliances, rising protectionism, and regional regulatory disparities disrupt once-stable global supply chains and business networks (Edler et al., 2023; Gopinath, 2023). Scholars now describe a multipolar, geoeconomically fragmented order in which competing blocs advance divergent technological standards and regulations, effectively creating rival digital spheres of influence. At the same time, digital technologies—cloud platforms, AI agents, blockchain-based ledgers, and immersive spatial media—continue to knit people, data, and devices together across borders at unprecedented speed and scale (Baiyere et al., 2023; Skog et al., 2018; Yoo et al., 2024). This creates a striking paradox: while economic and political landscapes fracture, digital ecosystems increasingly bridge these gaps.

Organizations and individuals now navigate a world in which fragmented geopolitics and diverging local regulations coexist with borderless, interoperable digital platforms that enable real-time data sharing, remote work and collaborative innovation (Bloom et al., 2024; Jovanovic et al., 2022; Yang et al., 2021). This push-and-pull dynamic creates both risk and opportunity: firms face inefficiencies and compliance hazards if they ignore fragmentation, yet digital integration opens new ways to coordinate across borders, optimize operations and cultivate global partnerships (Bockelmann et al., 2024). Given these dual forces, understanding how digital ecosystems can function as collaboration enablers in a fragmented world becomes paramount. Accordingly, the following guiding research question is articulated:

RQ: How can polycentric digital ecosystems serve as collaboration enablers in a fragmented, multipolar world?

To explore this tension, I focus on two core constructs: **fragmentation**—geopolitical, organizational, and technological—and **digital integration** through digital ecosystems. I contend that the prevailing platform ecosystem lens is too narrow to capture the complex interconnections that define today's digital world. Platform ecosystems are architecturally mono-centric constellations organized around a focal firm that owns the core, controls boundary resources and orchestrates multisided interactions (Parker et al., 2016).

By contrast, I define a digital ecosystem as *a polycentric socio-technical network of independent yet inter-dependent actors—individuals, organizations, devices, data services—that co-create and capture shared value through multiple, loosely coupled digital resource flows*. Platform ecosystems represent just one monocentric subtype within a much broader, ecological spectrum of digital connectedness.

Digital ecosystems are conceptualized as a **four-layer spectrum**: personal, organizational, inter-organizational, and global. These layers are seamlessly nested. Integration—and thus collaboration—within these ecosystems is enabled by a suite of technologies such as AI, blockchain, federated data spaces, and spatial computing. The unifying claim of this paper is that polycentric digital ecosystems can enable collaboration across all four layers and counter the centrifugal forces of geopolitical, organizational, and technological fragmentation. Methodologically, the framework was developed through a conceptual synthesis of literature in information systems, platform theory, and ecosystem governance.

This study makes three contributions to the Information Systems discourse. First, I reconceptualize the digital ecosystem construct by disentangling it from the dominant platform-ecosystem paradigm and developing a conceptual framework of polycentric digital ecosystems. This framework defines digital ecosystems as polycentric, multi-layered socio-technical systems that span personal, organizational, inter-organizational, and global domains, each with distinct governance logics but porous boundaries. Second, I identify the foundational technologies—AI automation, blockchain trust, federated data spaces, and immersive technologies—that enable the integration and functioning of digital ecosystems, and discuss how these technologies enable collaboration amid geopolitical, organizational, and technological fragmentation. Third, the framework is distilled into clear managerial and policy implications, outlining how governance, technology choices, and standards can be calibrated at each layer to foster digital integration and cross-border collaboration, thereby counteracting fragmentation.

The remainder of the paper proceeds as follows. Section 2 examines fragmentation in a multipolar world, outlining the geopolitical, organizational, and technological forces that challenge digital integration. Section 3 disentangles digital ecosystems from the dominant platform-ecosystem paradigm, tracing the literature's hub-centric bias and advancing a polycentric view that underpins the conceptual framework. Section 4 introduces the conceptual model, presenting the four-layer spectrum and identifying the foundational technologies that enable integration. Section 5 discusses managerial and policy implications, with particular emphasis on governance, trust, and security. The study concludes with a positive outlook on the role of digital ecosystems in fostering collaboration and collective intelligence in an increasingly complex world.

## Fragmentation in a Multipolar World

Global value chains now operate against a backdrop of what the IMF terms *geoeconomic fragmentation*—a multi-polar world characterized by rising geopolitical tensions, trade barriers, shifting alliances, and intensifying technological competition (Edler et al., 2023; Gopinath, 2023). At the center of these dynamics lies a contest for technological leadership, which increasingly defines both national power and global influence. The emergence of a multipolar world—characterized by multiple centers of power—complicates international cooperation. Unlike unipolar or bipolar systems, multipolarity introduces both opportunities and challenges, as nations navigate fluid alliances, decentralized decision-making, and diverse technological ecosystems. This study adopts a three-part lens:

- *Geopolitical fragmentation:* Evident in export controls, data-localization laws, and the strategic positioning of "connector" countries that straddle competing bloc (Aiyara & Ohnsorged, 2024).
- *Organizational fragmentation:* (a) Regulatory complexity: Organizations must adapt processes to comply with divergent local data and AI regulations.; (b) Operational inefficiencies: Challenges include restricted market access, supply chain disruptions, and fragmented local technology stacks. (c) Workforce dispersion: Organizations must manage remote and cross-border teams in an environment of shifting political conditions (Baptista et al., 2020; Bloom et al., 2024; Constantiou et al., 2023; Kellogg et al., 2020; Yang et al., 2021).
- *Technological fragmentation:* While legacy systems persist, they increasingly coexist with disruptive technologies. From quantum–AI integrations (Castelvecchi, 2024) to spatial computing platforms (Schmitt, 2023), emerging technologies promise new modes of connectivity—but also introduce fresh coordination challenges (Skog et al., 2018).

Together, these forces create a landscape of "default decentralization," in which cross-border collaboration depends not on assumed global cohesion, but on deliberate, digitally enabled integration to overcome the centrifugal pull of fragmentation.

## From Platform to Polycentric Digital Ecosystems

Mainstream platform-ecosystem theory is built around a central orchestrator that controls boundary resources and multisided interactions (Parker et al., 2016, 2017). IS work treats "digital ecosystem" largely as a synonym for "platform ecosystem," implicitly assuming a single focal core that orchestrates value creation. For example, Fraunhofer directly equates a digital ecosystem with a partner network around a digital platform. Governance, value capture and network-effects are viewed mainly from the vantage point of the firm that launches the platform.[1] Frauenhofer puts a single platform in the structural core. This is the same architectural assumption that underlies the platform-ecosystem literature. Kretschmer et al. (2022) reinforce the mono-centric view by framing the platform leader as the "visible hand" that coordinates the ecosystem and captures value. However, recent literature calls for a broader, ecological reading of digital ecosystems. Kapoor et al. (2021) echo this concern, urging a socio-technical perspective yet still conceptualize the ecosystem as a network centered on a single platform, reinforcing the hub-centered bias I seek to transcend. Hein et al. (2020) broaden the discussion by showing that platform ownership can range from single-firm to consortium to peer-to-peer, but even this extended typology remains hub-oriented and thus too narrow for today's polycentric digital architectures. In contrast, Senyo et al. (2019) define a digital-business ecosystem as an open, self-organizing socio-technical environment of loosely coupled actors, an explicitly polycentric view. These studies collectively underpin the spectrum framework: digital ecosystems are polycentric and span personal, organizational, inter-organizational, and global layers, each with distinct governance logics but porous boundaries.

*Digital ecosystem.* A socio-technical network of two or more independent yet inter-dependent actors (individuals, organizations, devices, data services) that co-create and capture a shared value proposition through multiple, loosely coupled, digitally mediated resource flows. Governance is distributed and adaptive.

*Platform ecosystem.* A subset of the above that is architecturally and contractually organized around one focal digital platform (or tightly coupled platform stack). Value is driven by multisided network effects centered on the platform's core-periphery structure.

Recognizing this distinction is critical for theorizing collaboration under fragmentation. Because platform ecosystems rely on a hub-and-spoke architecture, they may replicate geopolitical chokepoints; by contrast, polycentric digital ecosystems offer alternative topologies—data spaces, federated clouds, blockchain-enabled networks—that can reroute around political or technical barriers. Building on the literature review, I conceptualize digital ecosystems as a four-layer spectrum that scales from the individual to the planetary level (Figure 1).

## Digital Ecosystems as Collaboration Enablers

Digital ecosystems are the foundation of future collaboration in the age of AI. This section examines how they can serve as powerful enablers of cooperation in an increasingly fragmented world. It encapsulates the central thesis: Fragmentation (geopolitical, organizational, technological) is pulling actors apart, while digital integration, realized through layered digital ecosystems (personal, organizational, inter-organizational, global), pulls them back together. Integration is powered by four technology clusters—AI and automation, blockchain trust, federated data spaces, and immersive technologies. While several layered models have been proposed to conceptualize digital infrastructures, platforms, and ecosystems—often separating technical, organizational, or application layers (de Reuver et al., 2018; Oberländer et al., 2025; Yoo et al., 2010)—a complementary stream has examined governance within such ecosystems, typically from the perspective of platform owners or centralized coordination (Constantinides et al., 2018; Wareham et al., 2014). This study takes a different lens. The proposed spectrum framework goes beyond a technical architecture, positioning polycentric governance (Ostrom, 2010) as the organizing principle across the personal, organizational, inter-organizational, and global layers. Digital integration is achieved not merely by aligning technical components and data flows but by coordinating semi-independent actors around shared rules, interfaces, and incentives. The unit of analysis therefore shifts from platform-centric control and data pipelines to patterns of coordination and autonomy that enable collective intelligence to emerge across scales within fragmented environments.

---

[1] Digital Ecosystems and Digital Platforms: https://www.iese.fraunhofer.de/en/services/digital-ecosystems.html

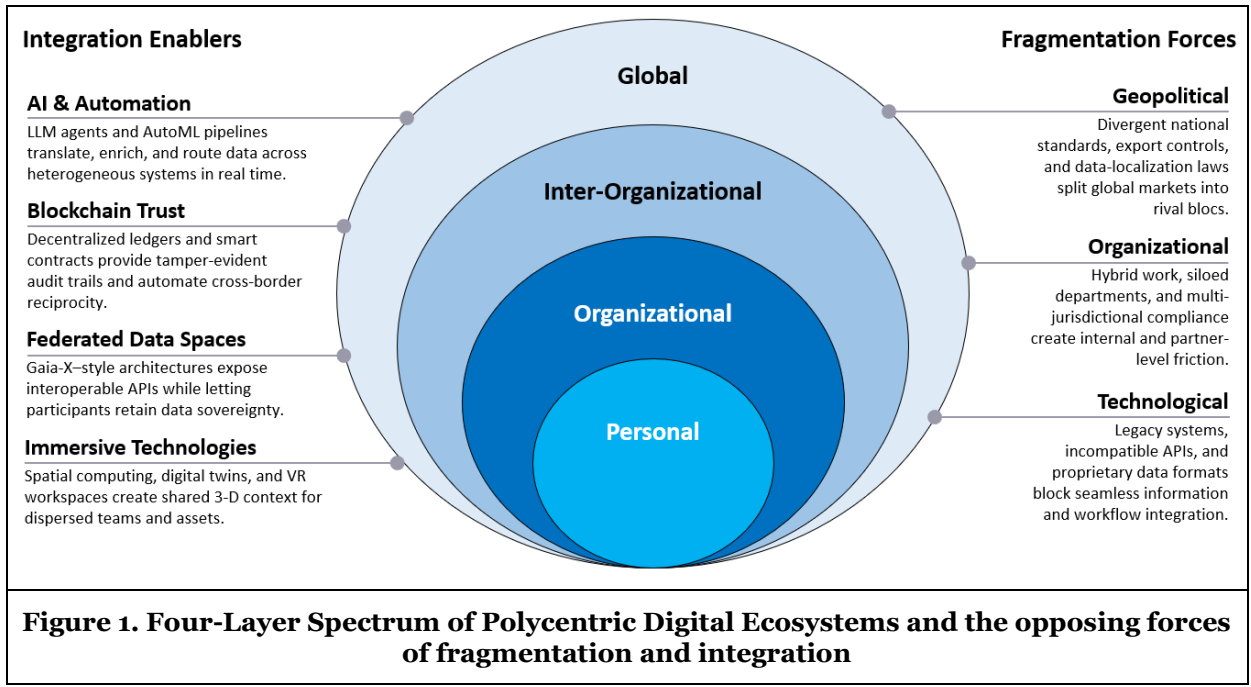

**Figure 1. Four-Layer Spectrum of Polycentric Digital Ecosystems and the opposing forces of fragmentation and integration**

## *The Spectrum of Digital Ecosystems*

Digital ecosystems are not a monolithic concept but can manifest in different sizes and configurations. To analyze how they counter geopolitical, organizational and technological fragmentation, I conceptualize digital ecosystems as a polycentric four-layer spectrum—personal, organizational, inter-organizational and global (Figure 1). They offer a path to digital integration, not as a one-time fix, but as a dynamic and scalable architecture that enables ongoing collaboration, knowledge sharing, and the emergence of collective intelligence across borders. The layers are nested and porous: a remote worker (personal layer) uses an enterprise Slack space (organizational layer), which feeds a sectoral data space (inter-organizational layer) running on the public Internet (global layer).

**Personal ecosystems**. These ecosystems involve single-user or small-team tools—such as cloud-based collaboration suites, personal project management apps, or communication platforms—often operating without strict corporate or geopolitical constraints. They are easy to adopt but rely heavily on user-centric security (e.g., passwords, multi-factor authentication). For instance, a freelance developer contributing to open-source projects via GitHub and coordinating asynchronously through Slack exemplifies how individuals operate autonomously yet collaboratively in loosely structured, polycentric environments.

**Organizational ecosystems**. At the enterprise level, digital ecosystems are formalized and secure. They unify departments through data lakes, AI-driven applications, and internal collaboration platforms—while employing robust security (VPNs, Zero Trust architectures, encryption). This ensures that the organization can harness collective intelligence internally, yet maintain control over sensitive data and IP. An illustrative example is a multinational enterprise in which semi-autonomous business units collaborate through a shared AI-enhanced knowledge platform—aligning decisions across the organization while preserving local control over data and operations.

**Inter-organizational ecosystems**. Here, multiple organizations (e.g., supply chain partners, industry consortia) collaborate via shared platforms, standardized APIs, and federated trust protocols. Each participant retains autonomy over its internal data but agrees on common data standards and compliance frameworks for secure information exchange. Catena-X, a federated automotive data space, illustrates this layer by enabling manufacturers and suppliers to share digital twin data securely, without centralizing ownership or control.

**Global ecosystems**. At the largest scale, ecosystems span international and cross-industry boundaries (e.g., global open-source communities, decentralized blockchain networks). They often rely on open standards and broad stakeholder participation. Although they enable enormous network effects and collective intelligence on a planetary scale, they also must navigate heterogeneous regulations, legal frameworks, and cultural norms. The global Internet itself serves as a clear example: a polycentric, standards-based infrastructure governed by multiple autonomous yet interdependent communities (e.g., open-source networks, DAOs, protocol-based platforms). Each operates under its own rules and governance mechanisms, yet they are integrated through shared protocols.

*Technological Enablers*

Digital ecosystem integration is not achieved by a single technology but by four complementary clusters that tackle different facets of fragmentation. The OECD 2024 Digital Economy Outlook highlights AI, immersive XR, high-speed networks, and federated data spaces as system-wide enablers, arguing that interoperable standards and inclusive governance are essential to translating these technologies into cross-border collaboration.[2] Building on this perspective, I identify four technology clusters that function as integration glue across the spectrum. The primary technologies that facilitate location-independent collaboration in a fragmented world are the following:

**AI & automation.** Large-language-model agents can already draft code, guide robotic design and triage medical images (Moor et al., 2023; Stella et al., 2023). By embedding reasoning into APIs, they dissolve interface mismatches and enable location-independent collaboration. AI will not only enable transformation and integration across a wide spectrum of industries (Jansen et al., 2025; Schmitt, 2025), but will be a primary building block of modern organizations. E.g. Microsoft introduced the concept of the frontier firm characterized by autonomous agent teams that are orchestrated by humans – the Agent Boss.

**Blockchain trust.** Bitcoin showed that secure, cross-border value exchange is possible without a central adjudicator (Nakamoto, 2008). Smart contracts extend this to *any* digital asset, allowing "co-operation among strangers" and mitigating the trust deficit that geoeconomic fragmentation creates (Gregory et al., 2024).

**Federated data spaces.** Unlike traditional data lakes, Gaia-X-style federations implement polycentric governance: every participant keeps its data under its own legal jurisdiction, yet subscribes to common identity, policy and semantics standards. This model directly addresses technological fragmentation and cross-border data-sovereignty barriers. Recent pilots (e.g., Catena-X) show how a federated control plane lets firms share digital-twin data without surrendering custody.

**Immersive technologies.** A further enabler of global collaboration is the increasing adoption of spatial technologies, such as virtual and augmented reality (Schmitt, 2023). The internet has laid the foundation for digital ecosystems. The Web gave us 2-D pages and links and access to global information. The mobile era then transplanted this functionality onto touchscreen smartphones freeing users from fixed locations. Now a spatial era is emerging, which is characterized by a persistent, real-time network of 3-D spaces and objects, which allows us to step inside, anchored to physical coordinates but rendered, updated, and shared through mixed-reality devices. These spatial ecosystems (also referred to as the Metaverse or Extended Reality Ecosystems) can be accessed via *spatial hardware* (e.g., the Vision Pro, Quest 4). The emerging spatial web—persistent 3-D spaces anchored to physical coordinates—lets engineers, surgeons or scientists share a *situated* digital twin of the artifact they co-create, accelerating sense-making across borders (Schmitt, 2023).

Taken together, these clusters form the technical stack that "glues" the four-layer spectrum, making it possible to move information, value and situational awareness fluidly from the personal edge to global infrastructures while maintaining security and trust.

---

[2] OECD Digital Economy Outlook 2024: https://www.oecd.org/en/publications/2024/05/oecd-digital-economy-outlook-2024-volume-1_d30a04c9.html

# Discussion

This study introduced a spectrum framework of *polycentric digital ecosystems*. Digital ecosystems are interconnected networks of technologies, platforms, and stakeholders that share and exchange data seamlessly. Rather than operating as siloed applications or organizational units, these ecosystems unify people, processes, and information in a common digital environment. By aligning disparate systems through open standards, APIs, and AI-driven integrations, they enable fluid collaboration across geographical, organizational, and functional boundaries. This deep connectivity fosters real-time knowledge sharing and amplifies innovation, as insights gained in one domain can be rapidly applied elsewhere. As businesses and societies grapple with increasingly decentralized work structures, digital ecosystems emerge as the foundational pillar that underpins secure, scalable, and inclusive collaboration in the age of AI.

## *Managerial and Policy Implications*

In today's connected world, individuals, organizations, and institutions no longer act in isolation. Instead, they operate as nodes within platform-based constellations where value is co-created and co-captured (Cennamo, 2019). Within this milieu, effective digital strategizing shifts from episodic IT planning to continuous, data-driven sense-and-respond cycles that align ecosystem participants around shared objectives (Morton et al., 2022).

Digital maturity, therefore, is not simply about acquiring the latest technologies alone, but about orchestrating complementary capabilities across permeable borders (Kane et al., 2015). Coordination mechanisms—such as APIs, AI agents, and modular governance protocols—function as connective tissue that synchronizes distributed actors in real time (Leong et al., 2024). These mechanisms unlock collaborative exploration, enabling AI-augmented teams to navigate expansive design and knowledge spaces faster and more effectively than any one actor could alone (Wang et al., 2023).

As innovation accelerates, platform leaders and ecosystem stewards must govern access, set interoperability standards, and calibrate incentives to harness emergent, network-driven opportunities while minimizing fragmentation risks (Kari et al., 2025; Quero et al., 2025). Strategy in digital ecosystems, then, becomes the art of designing adaptive, polycentric architectures that let intelligence, innovation, and trust flow across a diverse value network. Several design imperatives emerge:

*Layer-sensitive strategy*. Managers should tailor governance and security models to the specific layer in which they operate. Personal ecosystems thrive on usability and lightweight multi-factor authentication; inter-organizational data spaces require federated identity management, shared taxonomies, and legally binding SLAs; global ecosystems demand blockchain-grade auditability and multi-stakeholder oversight.

*Tech-stack orchestration*. Deploying multiple enablers in concert—such as AI agents that broker data in a Gaia-X federation and notarize transactions on a permissioned blockchain—generates a "stacking effect" that counters both technological and geopolitical fragmentation more effectively than siloed tools.

*Workforce dispersion*. Evidence on hybrid work (Bloom et al., 2024; Yang et al., 2021) shows that productivity and retention are not harmed when location independence is supported by robust digital ecosystems. Organizations should therefore treat immersive workspaces and AI co-pilots as mainstream infrastructure, not fringe benefits.

*Policy alignment*. Regulators aiming to balance data sovereignty with innovation can adopt the polycentric data-space model: mandate interoperability standards and audit trails, but allow data to remain under national or organizational control. Connector countries that embrace such frameworks can position themselves as trusted hubs in a fractured global economy.

These implications support the central claim that the purposeful design of polycentric, technology-enabled digital ecosystems can shift the balance from fragmentation toward integration. In doing so, they create resilient pathways for collective intelligence and innovation in an era defined by volatility, interdependence, and multipolar uncertainty.

*Limitations and Future Research*

This paper offers a conceptual framework for understanding how polycentric digital ecosystems can enable collaboration amid geopolitical, organizational, and technological fragmentation. While the framework is grounded in a synthesis of emerging literature and current practice, it remains theoretical in nature. Although the study naturally adopts an optimistic perspective on the integrative potential of polycentric digital ecosystems, it is important to recognize the broader structural limitations. Fragmentation is often rooted in entrenched political and economic dynamics. Even if polycentric architectures can facilitate collaboration, many digital ecosystems remain dependent on dominant platform infrastructures that exert significant technological and economic power, often in asymmetrical or even extractive ways. These structural constraints deserve further investigation and call for a more critical exploration of how genuine polycentricity can be sustained within existing global power structures. Future research should empirically investigate how the proposed four-layer spectrum manifests across different industries, regions, and governance contexts. Comparative case studies, design science research, or longitudinal fieldwork could enrich our understanding of how specific technologies—such as AI agents or spatial computing—operate as enablers in fragmented environments. As digital interdependence grows, cybersecurity will become increasingly central to maintaining trust, resilience, and operational continuity across ecosystem layers (Schmitt & Koutroumpis, 2025). In parallel, research on human–AI collaboration is needed to explore how distributed, AI-augmented teams interact, make decisions, and co-create value in polycentric digital environments.[3] As digital ecosystems continue to evolve, ongoing research will be essential to refine, adapt, and test the framework under varying conditions of volatility and complexity.

# Conclusion

As geopolitical, organizational, and technological fragmentation accelerates, the need for resilient, collaborative structures has never been greater. Against this backdrop, this paper has argued that polycentric digital ecosystems offer a viable pathway to foster collaboration in an increasingly fragmented, multipolar world. By disentangling the digital ecosystem construct from the dominant platform ecosystem paradigm, I developed a four-layer spectrum that captures the nested nature of digital interaction, from individual to global scale, and identified four technology clusters—AI and automation, blockchain trust, federated data spaces, and spatial computing—that enable integration across these layers. The framework contributes to the Information Systems discourse by illuminating how polycentric digital ecosystems can be purposefully orchestrated to counteract fragmentation, enable collaboration, and unlock collective intelligence.

# References

Aiyara, S., & Ohnsorged, F. (2024). Geoeconomic Fragmentation and "Connector" Countries.

Baiyere, A., Grover, V., Lyytinen, K. J., Woerner, S., & Gupta, A. (2023). Digital "x"—Charting a Path for Digital-Themed Research. Information Systems Research. https://doi.org/10.1287/isre.2022.1186

Baptista, J., Stein, M.-K., Klein, S., Watson-Manheim, M. B., & Lee, J. (2020). Digital work and organisational transformation: Emergent Digital/Human work configurations in modern organisations. The Journal of Strategic Information Systems, 29(2), 101618. https://doi.org/10.1016/j.jsis.2020.101618

Bloom, N., Han, R., & Liang, J. (2024). Hybrid working from home improves retention without damaging performance. Nature, 630(8018), 920–925. https://doi.org/10.1038/s41586-024-07500-2

Bockelmann, T., Werder, K., Recker, J., Lehmann, J., & Bendig, D. (2024). Configuring alliance portfolios for digital innovation. The Journal of Strategic Information Systems, 33(1), 101808. https://doi.org/10.1016/j.jsis.2023.101808

Castelvecchi, D. (2024). The AI–quantum computing mash-up: will it revolutionize science? Nature. https://doi.org/10.1038/d41586-023-04007-0

Cennamo, C. (2019). Competing in Digital Markets: A Platform-Based Perspective. Academy of Management Perspectives, 35(2), 265–291. https://doi.org/10.5465/amp.2016.0048

---

[3] https://www.microsoft.com/en-us/worklab/work-trend-index/2025-the-year-the-frontier-firm-is-born

Constantinides, P., Henfridsson, O., & Parker, G. G. (2018). Introduction—Platforms and Infrastructures in the Digital Age. Information Systems Research, 29(2), 381–400. https://doi.org/10.1287/isre.2018.0794

Constantiou, I., Joshi, M., & Stelmaszak, M. (2023). Organizations as Digital Enactment Systems: A Theory of Replacement of Humans by Digital Technologies in Organizational Scanning, Interpretation, and Learning. Journal of the Association for Information Systems, 24(6), 1770–1798. https://doi.org/10.17705/1jais.00833

de Reuver, M., Sørensen, C., & Basole, R. C. (2018). The Digital Platform: A Research Agenda. Journal of Information Technology, 33(2), 124–135. https://doi.org/10.1057/s41265-016-0033-3

Edler, J., Blind, K., Kroll, H., & Schubert, T. (2023). Technology sovereignty as an emerging frame for innovation policy. Defining rationales, ends and means. Research Policy, 52(6), 104765. https://doi.org/10.1016/j.respol.2023.104765

Gopinath, G. (2023). Cold War II? Preserving Economic Cooperation Amid Geoeconomic Fragmentation.

Gregory, R. W., Beck, R., Henfridsson, O., & Yaraghi, N. (2024). Cooperation Among Strangers: Algorithmic Enforcement of Reciprocal Exchange with Blockchain-Based Smart Contracts. Academy of Management Review. https://doi.org/10.5465/amr.2023.0023

Hein, A., Schreieck, M., Riasanow, T., Setzke, D. S., Wiesche, M., Böhm, M., & Krcmar, H. (2020). Digital platform ecosystems. Electronic Markets, 30(1). https://doi.org/10.1007/s12525-019-00377-4

Jansen, M., Nguyen, H. Q., & Shams, A. (2025). Rise of the Machines: The Impact of Automated Underwriting. Management Science, 71(2), 955–975. https://doi.org/10.1287/mnsc.2024.4986

Jovanovic, M., Sjödin, D., & Parida, V. (2022). Co-evolution of platform architecture, platform services, and platform governance: Expanding the platform value of industrial digital platforms. Technovation, 118, 102218. https://doi.org/10.1016/j.technovation.2020.102218

Kane, G. C., Palmer, D., Phillips, A. N., Kiron, D., & Buckley, N. (2015). Strategy, not Technology, Drives Digital Transformation: Becoming a digitally mature enterprise. https://doi.org/10.1176/appi.ajp.159.9.1620

Kapoor, K., Ziaee Bigdeli, A., Dwivedi, Y. K., Schroeder, A., Beltagui, A., & Baines, T. (2021). A socio-technical view of platform ecosystems: Systematic review and research agenda. Journal of Business Research, 128. https://doi.org/10.1016/j.jbusres.2021.01.060

Kari, A., Bellin, P., Matzner, M., & Gersch, M. (2025). Governing the emergence of network-driven platform ecosystems. Electronic Markets, 35(1), 4. https://doi.org/10.1007/s12525-024-00745-9

Kellogg, K. C., Valentine, M. A., & Christin, A. (2020). Algorithms at Work: The New Contested Terrain of Control. Academy of Management Annals, 14(1), 366–410. https://doi.org/10.5465/annals.2018.0174

Kretschmer, T., Leiponen, A., Schilling, M., & Vasudeva, G. (2022). Platform ecosystems as meta-organizations: Implications for platform strategies. Strategic Management Journal, 43(3). https://doi.org/10.1002/smj.3250

Leong, C., Lin, S., Tan, F., & Yu, J. (2024). Coordination in a Digital Platform Organization. Information Systems Research, 35(1), 363–393. https://doi.org/10.1287/isre.2023.1226

Moor, M., Banerjee, O., Abad, Z. S. H., Krumholz, H. M., Leskovec, J., Topol, E. J., & Rajpurkar, P. (2023). Foundation models for generalist medical artificial intelligence. Nature, 616(7956), 259–265. https://doi.org/10.1038/s41586-023-05881-4

Morton, J., Amrollahi, A., & Wilson, A. D. (2022). Digital strategizing: An assessing review, definition, and research agenda. The Journal of Strategic Information Systems, 31(2), 101720. https://doi.org/10.1016/j.jsis.2022.101720

Nakamoto, S. (2008). Bitcoin: A peer-to-peer electronic cash system. Satoshi Nakamoto.

Oberländer, A. M., Karnebogen, P., Rövekamp, P., Röglinger, M., & Leidner, D. E. (2025). Understanding the influence of digital ecosystems on digital transformation: The OCO (orientation, cooperation, orchestration) theory. Information Systems Journal, 35(1), 368–413. https://doi.org/10.1111/isj.12539

Ostrom, E. (2010). Beyond Markets and States: Polycentric Governance of Complex Economic Systems. American Economic Review, 100(3), 641–672. https://doi.org/10.1257/aer.100.3.641

Parker, G., Van Alstyne, M., & Jiang, X. (2017). Platform ecosystems: How developers invert the firm. MIS Quarterly: Management Information Systems, 41(1). https://doi.org/10.25300/MISQ/2017/41.1.13

Parker, G., van Alstyne, M. W., & Choudary, S. P. (2016). Platform Revolution: How Networked Markets are Transforming the Economy and How to Make Them Work for You. W.W.Norton & Company.

Quero, M. J., Díaz-Méndez, M., & Ruiz-Alba, J. L. (2025). How does innovation emerge in open platform ecosystems? Electronic Markets, 35(1), 8. https://doi.org/10.1007/s12525-024-00749-5

Schmitt, M. (2023). Metaverse: Implications for Business, Politics, and Society. SSRN Electronic Journal. https://doi.org/10.2139/ssrn.4168458

Schmitt, M. (2025). Leveraging AutoML and Explainable AI for Digital Transformation. In Achieving Digital Transformation through Analytics and AI (pp. 211–229). World Scientific. https://doi.org/10.1142/9789811296475_0011

Schmitt, M., & Koutroumpis, P. (2025). Cyber Shadows: Neutralizing Security Threats with AI and Targeted Policy Measures. IEEE Transactions on Artificial Intelligence, 1–9. https://doi.org/10.1109/TAI.2025.3527398

Senyo, P. K., Liu, K., & Effah, J. (2019). Digital business ecosystem: Literature review and a framework for future research. International Journal of Information Management, 47, 52–64. https://doi.org/10.1016/j.ijinfomgt.2019.01.002

Skog, D. A., Wimelius, H., & Sandberg, J. (2018). Digital Disruption. Business & Information Systems Engineering, 60(5), 431–437. https://doi.org/10.1007/s12599-018-0550-4

Stella, F., Della Santina, C., & Hughes, J. (2023). How can LLMs transform the robotic design process? Nature Machine Intelligence, 5(6), 561–564. https://doi.org/10.1038/s42256-023-00669-7

Wang, H., Fu, T., Du, Y., Gao, W., Huang, K., Liu, Z., Chandak, P., Liu, S., Van Katwyk, P., Deac, A., Anandkumar, A., Bergen, K., Gomes, C. P., Ho, S., Kohli, P., Lasenby, J., Leskovec, J., Liu, T.-Y., Manrai, A., … Zitnik, M. (2023). Scientific discovery in the age of artificial intelligence. Nature, 620(7972), 47–60. https://doi.org/10.1038/s41586-023-06221-2

Wareham, J., Fox, P. B., & Cano Giner, J. L. (2014). Technology Ecosystem Governance. Organization Science, 25(4), 1195–1215. https://doi.org/10.1287/orsc.2014.0895

Yang, L., Holtz, D., Jaffe, S., Suri, S., Sinha, S., Weston, J., Joyce, C., Shah, N., Sherman, K., Hecht, B., & Teevan, J. (2021). The effects of remote work on collaboration among information workers. Nature Human Behaviour, 6(1), 43–54. https://doi.org/10.1038/s41562-021-01196-4

Yoo, Y., Henfridsson, O., Kallinikos, J., Gregory, R., Burtch, G., Chatterjee, S., & Sarker, S. (2024). The Next Frontiers of Digital Innovation Research. Information Systems Research, 35(4), 1507–1523. https://doi.org/10.1287/isre.2024.editorial.v35.n4

Yoo, Y., Henfridsson, O., & Lyytinen, K. (2010). The new organizing logic of digital innovation: An agenda for information systems research. Information Systems Research, 21(4), 724–735. https://doi.org/10.1287/isre.1100.0322